\documentclass{kapproc} 

\usepackage{T1enc}
\usepackage{latexsym}
\usepackage{graphicx}
\usepackage{procps}

\begin{document}

\articletitle{OBSERVATIONS OF COOLING NEUTRON STARS}

\author{J. E. TR\"UMPER}

\affil{Max-Planck-Institut\\
f\"{u}r extraterrestrische Physik\\
85741 Garching\\
Germany}

\begin{abstract}
Observations of cooling neutron stars allow to measure 
photospheric radii and to constrain the equation of state of nuclear matter 
at high densities. In this paper we concentrate on neutron stars, which show 
thermal (photospheric) X-ray emission and have measured distances. After a 
short summary of the radio pulsars falling into this category we review the 
observational data of the 7 radio quiet isolated neutron stars discovered by 
ROSAT which have been studied in detail by Chandra, XMM-Newton and optical 
observations. Their spectra show blackbody temperatures between 0.5 and 1 
million Kelvin and an optical excess of a factor of 5-10 over the 
extrapolation of the X-ray spectrum. Four of these sources show 
periodicities between 3.45 and 11.37\,sec indicating slow rotation. The 
pulsed fractions are small, between 6 and 18 {\%}. The magnetic fields 
derived from spin down and/or possible proton cyclotron lines are of the 
order 10$^{13}$-10$^{14}$ G. We then discuss RX\,J1856.5--3754 in detail and 
suggest that the remarkable absence of any line features in its X-ray 
spectrum is due to effects of strong magnetic fields ($\sim $10$^{13}$ G). 
Assuming blackbody emission to fit the optical and X-ray spectrum we derive 
a conservative lower limit of the ``apparent'' neutron star radius of 
16.5\,km\,$\times$\,(d/117 pc). This corresponds to the radius for the ``true'' 
radius of 14\,km for a 1.4\,M$_{\odot}$ neutron star, indicating a stiff equation 
of state at high densities. A comparison of the result with mass-radius 
relations shows that in this case a quark star or a neutron star with a 
quark matter core can be ruled out with high confidence.
\end{abstract}


\section*{Introduction and History}

1960's and 1970's: The rocket experiments and the early satellites like 
Uhuru and Ariel-5 were not sensitive enough to detect the weak and soft 
thermal emission of neutron stars. A speculation by Chiu (1964) that the 
X-rays from the Crab nebula were due to a hot neutron star with 
kT$\sim$4\,keV was soon disproved by the famous NRL lunar occultation experiment which 
found only an extended source (Bowyer et al. 1964).

1980's: The Einstein observatory gave the first sensitive upper limit for 
the temperature of the Crab pulsar, kT$<$0.2\,keV (3$\sigma )$ (Harnden \& 
Seward, 1984). It is remarkable the most recent upper limit obtained with 
Chandra is not much lower, namely $<$0.17\,keV (3$\sigma )$ (Weisskopf 2004). 
The obstacles are the huge magnetospheric emission and the large 
interstellar absorption, apart from problems with the Chandra-HRC timing. As 
far as other pulsars are concerned Einstein and EXOSAT yielded only upper 
limits for their thermal emission as well.

1990's: A breakthrough came with ROSAT due to the excellent soft response of 
its PSPC. Thermal emission from a number of pulsars could be clearly 
identified while ASCA measured the ``hard power law tails'' which are of 
magnetospheric origin. Among these sources are PSR 1055--52, PSR 0656+14 and 
the newly discovered Geminga which have been called the ``three 
musceteers''. Perhaps even more important was the ROSAT discovery of a new 
class of thermally emitting neutron stars, which show no radio emission and 
no hard spectral tails, viz. no indication for magnetospheric emission. 
These objects called ``isolated neutron stars'' are the main subject of this 
paper (Sometimes they have been called X-ray dim 
isolated neutron stars -- XDINS --, but this is misleading 
because they are quite bright in X-rays, but dim in optical light).

2000's: Recently radio pulsars and isolated neutron stars have been studied 
extensively with the new powerful X-ray observatories Chandra and XMM-Newton 
which, taken together, provide a very substantial increase in collecting 
power, angular resolution, spectral bandwidth and spectral resolution 
compared with ROSAT and other previous missions.

One of the fundamental problems of neutron star physics is to determine the 
equation of state at supra-nuclear densities. In order to get a handle on 
that one must constrain the mass-radius relation and this can be done in 
principle by various methods which all have their specific problems: 
\begin{itemize}
\item Measurement of the gravitational redshift of spectral features. Problems: 
Identification of the feature, large spectral shifts in superstrong magnetic 
fields. 
\item Measurement of the surface gravity by analysing the radiative 
transfer in the neutron star photosphere. Problem: The method is not very 
sensitive and accurate. 
\item Measurement of characteristic frequencies (QPO) 
in accreting sources (see contributions of M.van der Klis, and F. Lamb in 
this volume). 
\item Measurement of the photospheric radius. Main problem: 
Requires know\-ledge of the source distance. 
\end{itemize}
We will discuss this method in 
more detail in this paper which will be organized as follows: In section 1 
we will give a short summary of the X-ray emitting radio pulsars 
concentrating on those sources which have measured parallaxes. In section 2 
we will review the properties of radio quiet isolated neutron stars, and 
section 3 will be devoted to the brightest of these sources, the enigmatic 
object RX\,J1856--3754 for which measurements of the parallax exist.

\section{X-ray emitting radio pulsars}

Three different X-ray spectral components have been identified in radio 
pulsars:
\begin{itemize}
\item The magnetospheric radiation, which is characterised by beaming and a 
power law spectrum. This component dominates the emission of very energetic 
pulsars and decreases rapidly with age.
\item Thermal emission from the polar caps, which are heated by the bombardment 
by high energy radiation/particles from the magnetosphere, or by heat 
outflow from the core region. This component has been detected in middle age 
pulsars like PSR\,0656+14 and millisecond pulsars (RX\,J0437--47). Since 
millisecond pulsars should have a cool core, their polar caps must be heated 
by magnetospheric bombardment.
\item Thermal radiation from the bulk surface, which is heated by core cooling. 
The alternative possibility of heating by low level accretion has not been 
positively identified yet.
\end{itemize}
What has just been said is reflected in the distribution of the $\sim$50 
X-ray detected radio pulsars in the P-dP/dt diagram shown in Fig.~1. 
Evidently, we see just those pulsars in X-rays, which have the largest spin 
down power ($\sim$dP/dt\,P$^{-3})$, or youngest age ($\sim$dP/dt\,P$^{-1})$.

The multicomponent spectra of three middle age pulsars depicted in Fig.~2 
clearly show the huge thermal peaks above the background of the broad-band 
nonthermal (magnetospheric) emissions. We note that a similar spectrum has 
been found in PSR\,1055--52, the third of the three musketeers. On the other 
hand, the rather young pulsar PSR\,J1811--1926 (64\,ms, 24000\,yrs) shows only 
a thermal component but no hard tail, which would indicate magnetospheric 
emission (Mc Gowan 2003). Obviously, the visibility of the beamed 
magnetospheric emission depends crucially on the orientation of spin axis 
and magnetic axis with respect to the line of sight.

Of special interest is the small subsample of pulsars for which distances 
are known from optical or radio parallaxes:

\begin{figure}[htbp]
\centerline{\includegraphics[width=4.60in,height=5.17in]{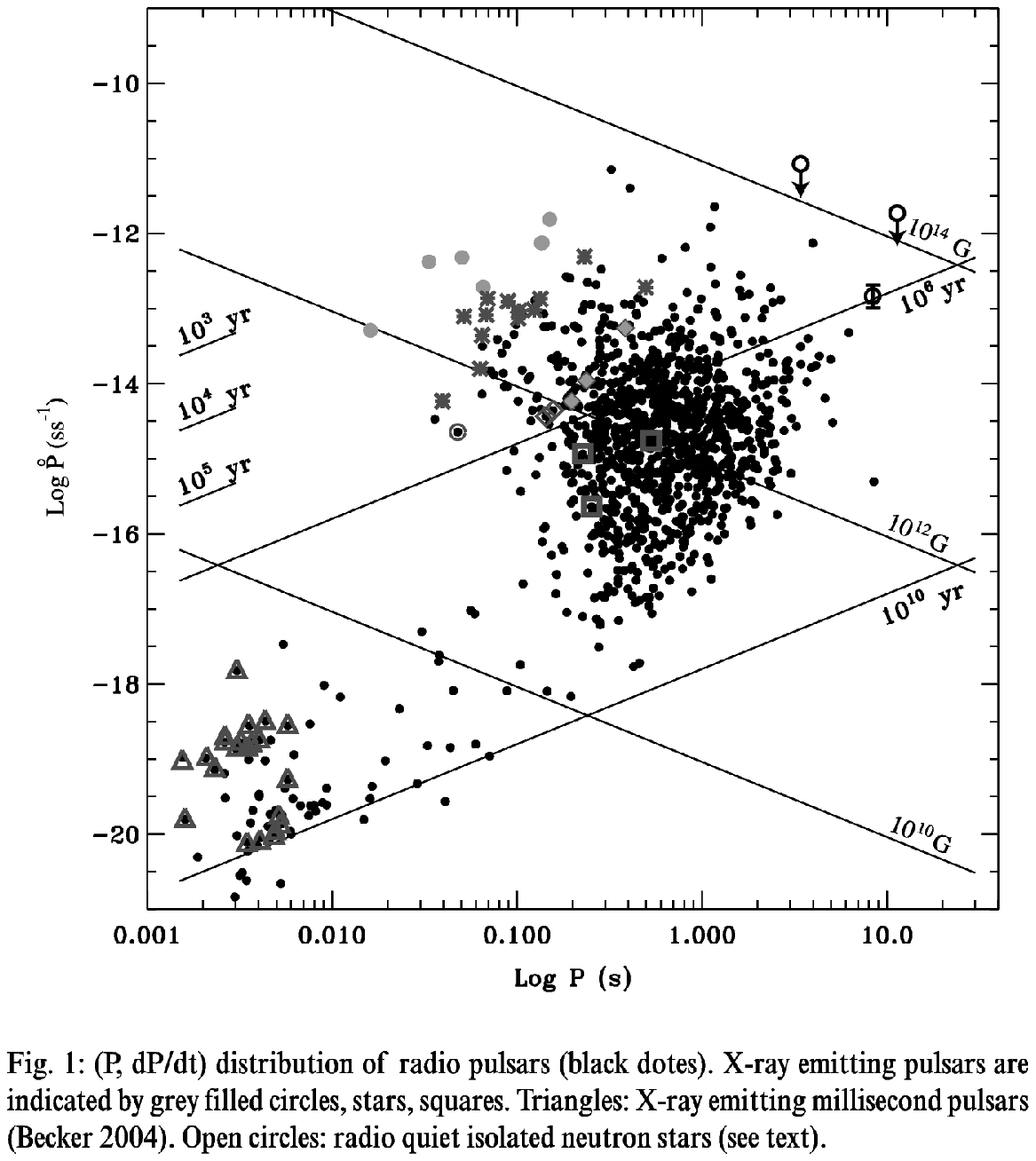}}
\label{fig1}
\end{figure}

For PSR 0656+14 the distance has been determined using the VLBA (Brisken et 
al. 2003), resulting in d\,=\,288$^{+33}_{-27}$\,pc which is significantly lower 
than the long used dispersion distance of 850pc. Using magnetized hydrogen 
atmospheric model fits Brisken et al. (2003) find a radius R$\sim $13-20 km, 
while the analysis of Pavlov et al. (2002) yields R$\sim$30\,km (scaled to 
the distance of 288\,pc). Here and in the following R represents the radius 
measured by a distant observer.

For Geminga the parallax has been determined from HST observations which 
give d\,=\,157$^{+59}_{-34}$\,pc and a blackbody radius R$_{bb}$\,$\sim$\,9\,km 
(Caraveo et al. 1996). According to Zavlin \& Pavlov (2002) the 
hydrogen/helium photospheric model fits in general yield radii which are 
larger by a factor 2-7 compared with blackbody fits, while those for 
magnetized H/He atmospheres are between nonmagnetic and blackbody radii. In 
the absence of a more detailed analysis of the soft Geminga spectrum we 
conclude that the radius is probably substantially larger than 10 km. 

\begin{figure}[htbp]
\centerline{\includegraphics[width=4.60in,height=3.91in]{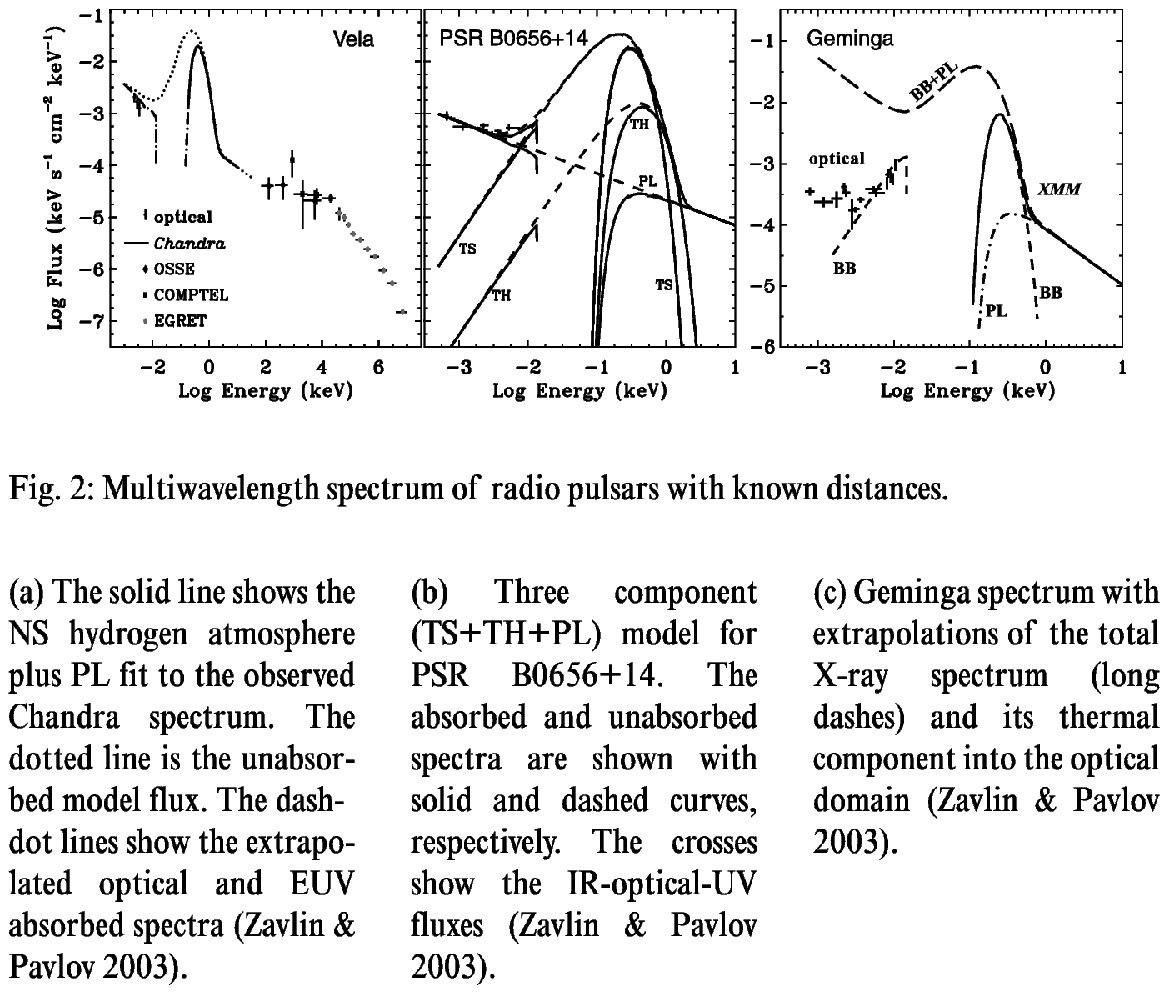}}
\label{fig2}
\end{figure}

The Vela pulsar has a VLB distance of 290\,pc (Dodson et al. 2003) which for a 
magnetized hydrogen atmosphere model leads to a radius R\,=\,17-20\,km (Pavlov 
et al. 2002).

In summary, the observations point to radii measured at infinity of 14-20\,km.

\section{Radio Quiet Isolated Neutron Stars }

The excellent soft X-ray response of ROSAT pointed and all-sky observations 
(see Tr\"{u}mper 1983) have led to the discovery of seven radio quiet, 
thermally emitting neutron stars which have been dubbed ``the magnificent 
seven''. A review of the observations can be found in a 
recent paper of Haberl 2004. The measured blackbody temperatures are between 
kT\,=\,44$-$96 eV, and the optical magnitudes are fainter than 24$^{m}$ (c.f. 
Table 1). All X-ray spectra show a low interstellar absorption, indicating 
that the objects are nearby, closer than a few hundred pc. For three of the 
sources proper motions could be measured which turn out to be very high, 
implying high space velocities and close distances. None of the sources 
shows an obvious association with a known supernova remnant, which suggests 
that they have ages $\ge$10$^{5}$\,yrs.

Four of the seven sources show X-ray pulsations with periods of typically 
10\,sec and pulsed fractions of typically 10{\%} (c.f. Table~1), suggesting that 
the neutron stars have an inhomogeneous temperature distribution.


\begin{table}[htbp]
\label{tab1}
\caption{Parameters of radio quiet isolated neutron stars (updated version
of Haberl 2004)}
\begin{center}
\begin{tabular}{lccccccc}
\hline
Source             &  P  & p. fr. &    L$_{x}$     & kT$_{BB}$ &  d   & Opt.  & pr. mot.\\
 name              & (s) & ({\%})   & (erg s$^{-1})$ & (eV)      & (pc) & (mag) & (mas/y)   \\
\hline
RX\,J0420.0-5022$^a$            &  3.45 & 12 & 2.7x10$^{30}$ & 44 & 100$^a$ & B$>$25.5         &  -  \\
RX\,J0720.4-3125            &  8.39 & 11 & 2.6x10$^{31}$ & 85 & 100$^a$ & B=26.6           &  97 \\
RX\,J0806.4-4123            & 11.37 &  6 & 5.7x10$^{30}$ & 95 & 100$^a$ & B$>$24           &  -  \\
1RXS\,J130848.6             & 10.31 & 18 & 5.1x10$^{30}$ & 90 & 100$^a$ & m$_{50CCD}$= &  -  \\
\multicolumn{1}{r}{+212708} &       &    &               &    &      &        28.6          &     \\          
RX\,J1605.3+3249            &   -   &  - & 1.1x10$^{31}$ & 92 & 100$^a$ & B$>$27           & 145 \\
RX\,J1856.5-3754            &   -   &  - & 1.5x10$^{31}$ & 63 & 117   & V=25.7          & 332 \\
1RXS\,J214303.7             &   -   &  - & 1.1x10$^{31}$ & 90 & 100$^a$ & R$>$23           &  -  \\
\multicolumn{1}{r}{+065419} &       &    &               &    &      &                  &     \\          
\hline
\end{tabular}

\end{center}
$^a$Assumed distance 
\end{table}

The slow down rate measured for RX\,J0720--3125 (henceforth RX\,J0720) leads to 
estimates of the magnetic field of $\sim$3\,$\times$\,10$^{12}$\,G and of the age of 
10$^{6}$\,years. A few of these objects exhibit small but significant changes 
of the spectra with pulse phase which may be explained by the anisotropic 
emission of strongly magnetized plasmas. Four of the sources show broad 
absorption line features which have been attributed to proton cyclotron 
absorption/scattering in magnetic fields of a few times 10$^{13}$ G (Table~2). 
In RX\,J0720 long term spectral changes have been found (de Vries et al. 
2004), which have been interpreted in terms of neutron star precession.

\begin{table}[htbp]
\label{tab2}
\caption{Magnetic field estimates for radio quiet isolated neutron stars 
(Haberl 2004)}
\begin{center}
\begin{tabular}{lccccc}
\hline
Object &  P  &       dP/dt           & E$_{cyc}$ & B$_{db}$      & B$_{cyc}$     \\
(name) & (s) & (10$^{-13}$ss$^{-1})$ & (eV)      & (10$^{-13}$G) & (10$^{-13}$G) \\
\hline
RX\,J0420.0-5022       & 3.45  &      $<92$     &   329   &  $<18$  & 6.6 \\
RX\,J0720.4-3125       & 8.39  & (1.4$\pm $0.6) &   262   & 2.8-4.2 & 5.2 \\
RX\,J0806.4-4123       & 11.37 &      $<18$     &    -    &  $<14$  &  -  \\
1RXS\,J130848.6+212708 & 10.31 &        -       & 100-300 &     -   & 2-6 \\
RX\,J1605.3+3249       &   -   &        -       & 450-480 &     -   & 9.1-9.7 \\
RX\,J1856.5-3754       &   -   &        -       &    -    & $\sim$1 &  -  \\
1RXS\,J214303.7+065419 &   -   &        -       &    -    &     -   &  -  \\
\hline
\end{tabular}

\end{center}
\end{table}

In summary, these findings strongly suggest that these ``magnificent seven'' 
are strongly magnetized (10$^{13}$-10$^{14}$G), slowly rotating neutron 
stars having an inhomogeneous temperature distribution over the stellar 
surface. Their main energy source must be heat loss from the hot interior 
(cooling), since accretion of matter from the interstellar medium is too 
inefficient due to the high stellar velocities. These sources do not show 
radio emission, probably because either they are evolved beyond the pulsar 
death line or because their radio beam is too narrow due to their large 
light cylinder radius.

\section{RX\,J1856--3754 }

\subsection{General Properties}

Among the radio quiet isolated neutron stars RX\,J1856.5-3754 (henceforth 
RX\,J1856) is the brightest and the only one with a known distance. Therefore 
it is best qualified for detailed studies aiming at a determination of its 
radius, and in the rest of this paper we concentrate primarily on this 
object.

RX\,J1856 was discovered serendipitously in a ROSAT PSPC field by Walter et 
al. (1996). Using the (HST) Walter {\&} Matthews (1997) identified the X-ray 
source with a faint blue star (V $\sim $ 26 mag). Its distance and proper 
motion were determined with the HST by Walter {\&} Lattimer (2002), to be 
(117$\pm$12) pc and 0.33 arcsec/year, respectively. With the VLT van 
Kerkwijk {\&} Kulkarni (2001) found a faint nebula surrounding the point 
source which has a cometary-like geometry with a 25? tail extending along 
the direction of motion. None of the X-ray observations revealed any 
variability on time scales up to ten years. The so far best upper limit of 
1.3{\%} (2$\sigma)$ on periodic variations in the range 10$^{-3}$--50\,Hz 
has been established by Burwitz et al. (2003) using a XMM-Newton EPIC-pn 
observation. Chandra LETG observations with high spectral resolution show a 
spectrum that can be fit by a Planckian spectrum with a temperature of 
63$\pm$3\,eV, (c.f. Fig.~3). Despite the excellent photon statistics and the 
good energy resolution of the LETG this spectrum is devoid of any spectral 
features. Compared with the optical spectrum which shows a Rayleigh-Jeans 
slope ($\sim\nu^{2})$, the X-ray spectrum is reduced by a factor of 
$\sim$6. Therefore, the overall spectrum of the source has often been 
described by a two-temperature blackbody model (e.g. Pons et al. 2002, 
Burwitz et al. 2003, Pavlov {\&} Zavlin 2003, Tr\"{u}mper et al. 2004).

\begin{figure}[htbp]
\centerline{\includegraphics[width=4.60in,height=4.19in]{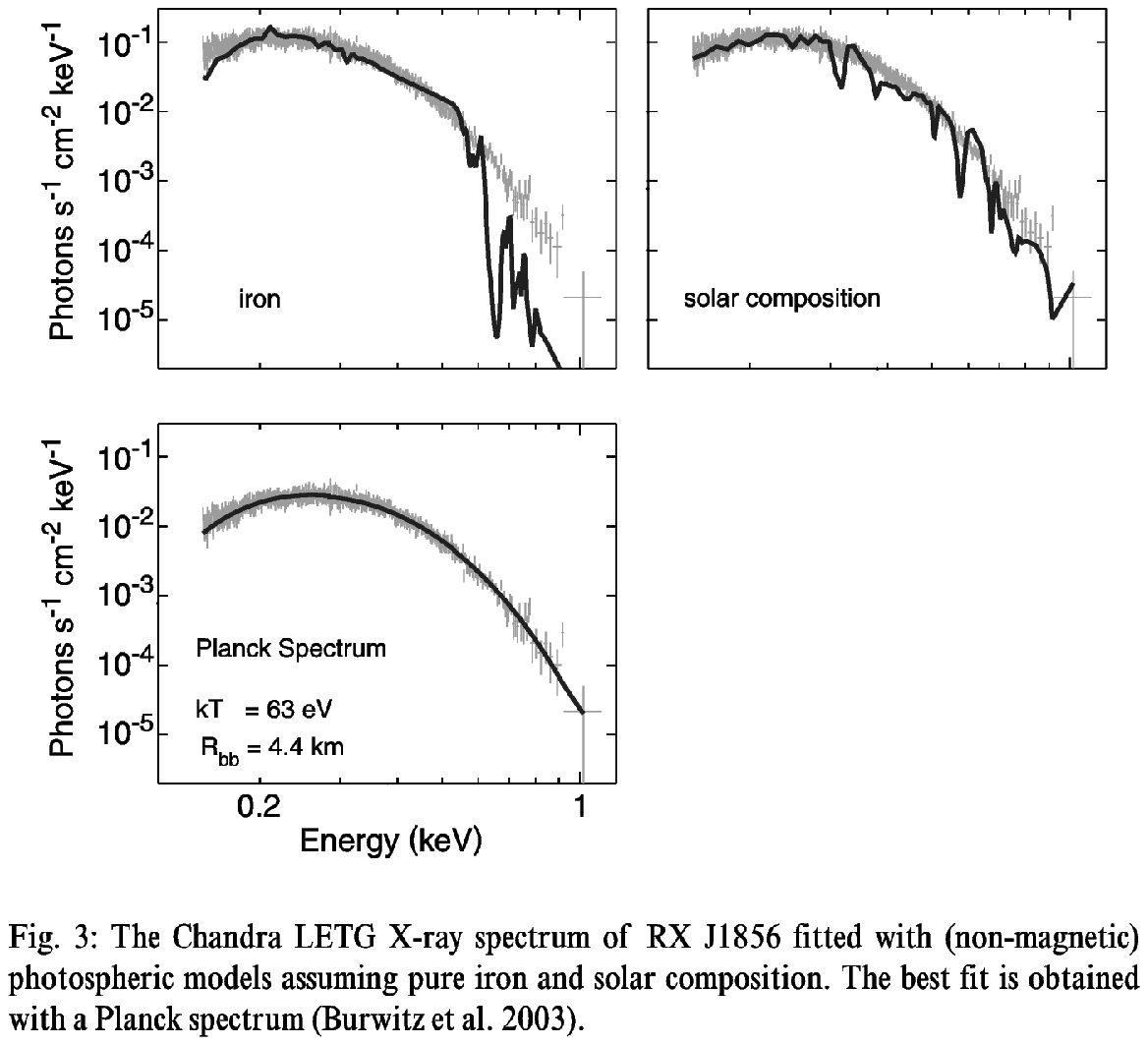}}
\label{fig3}
\end{figure}

A large number of papers have been dealing with the questions concerning the 
nature of this compact object and the proposed answers include everything 
from ``normal'' neutron stars with stiff or soft equations of state over 
neutron stars having a quarks core to bare (strange) quark stars, P-stars 
etc (for references c.f. Turolla et al. 2003). Before coming back to this 
topic we want to summarize some more observational data and their immediate 
consequences in somewhat more detail.

\subsection{The Magnetic Field Strength of RX\,J1856.5--3754}

The impressive lack of any significant spectral features in the LETG 
spectrum excludes magnetic fields of (1.3 - 7) $\times $ 10$^{11}$ G 
(electron cyclotron lines) and (2 - 13) $\times $ 10$^{13}$ G $^{ }$(proton 
cyclotron lines), see Burwitz et al. (2003). This leaves the possibility 
open of a low magnetic field characteristic for millisecond pulsars or a 
high magnetic field typical for normal pulsars. Unfortunately, due to the 
absence of a periodicity the usual estimate of the magnetic field of RX 
J1856 based on the rotating dipole model is not possible. Using 
phenomenological arguments based on the very small pulsed fraction in X-rays 
and on a comparison with other objects van Kerkwijk {\&} Kulkarni (2001)] 
have argued that the star has a relatively low magnetic field of a few 
10$^{11}$ G which may be marginally consistent with the absence of proton 
cyclotron lines. But this is not the only possibility. We estimate the 
magnetic field using the spin-down luminosity dE/dt $\sim $ 4 $\times $ 
10$^{32}$ erg/s required for powering the cometary-like emission nebula 
(Kerkwijk {\&} Kulkarni 2001) and the age of the star (t $\sim $ 5 $\times $ 
10$^{5}$ years) inferred from its proper motion and the distance to its 
likely birthplace in the Upper Sco OB association (Walter {\&} Lattimer 
2002). Applying the model of magnetic dipole braking we find a period of a 
$\sim $1.8\,sec and a magnetic field strength of $\sim $1.1 $\times $ 
10$^{13}$ G. We emphasise that these figures are very similar to those of 
the second brightest object of this kind, the pulsating source RX\,J0720 
whose spectral characteristics are very similar to those of RX\,J1856. While 
the estimate of dE/dt may be considered as rather reliable, the age derived 
from the birthplace argument is not so certain. However, an age of t $\sim $ 
5 $\times $ 10$^{5}$ years (with an uncertainty of a factor of two) is fully 
consistent with what we know empirically about the cooling of neutron stars. 
We therefore conclude that the magnetic field of RX\,J1856 is probably large, 
i.e. of the order of $>$10$^{13}$\,G. To confirm this, it is necessary to 
exclude the alternative hypothesis of a millisecond pulsar (van Kerkwijk 
\& Kulkarni 2001, Pavlov \& Zavlin 2003). To this end a high time 
resolution observation with XMM-Newton has already been scheduled.

\subsection{The Featureless X-ray Spectrum  of RX\,J1856.5--3754}

The main puzzle of RX\,J1856 is the observational fact that its X-ray 
spectrum (Fig.~3) is completely featureless. It has been pointed out by 
Burwitz et al. (2001, 2003) that nonmagnetic photospheric spectra assuming a 
pure iron composition are incompatible with the measured spectrum because 
the predicted Fe-L features are not detected with high significance. Even a 
solar composition model with its small abundance of metals leads to 
unacceptable spectral fits. Doppler smearing of the spectral lines due to 
fast rotation does not wash away completely the strongest spectral features 
(Braje {\&} Romani 2002, Pavlov et al. 2002)]. On the other hand hydrogen or 
helium photospheres can be excluded, because they would over-predict the 
optical flux by a very large factor (Pavlov et al. 1996). Therefore any 
nonmagnetic photosphere can be firmly excluded. This argument can be 
extended to magnetized hydrogen and helium photospheres (Zavlin {\&} Pavlov 
2002).

Iron photospheric models have been calculated by Rajagopal et al. 1997 for 
B=10$^{12.5}$ and 10$^{13}$G.$^{ }$Unfortunately they suffer from the fact 
that the radiative properties of iron atoms/ions in super strong magnetic 
fields are not known exactly, but only in Hartree-Fock approximation (work 
of Neuhauser et al. (1986). The resulting spectra contain a lot of lines 
having spacings of 50-100 eV, which could be easily resolved by the LETG 
(resolution $<$1 eV). However, according to Neuhauser et al. (1987) the iron 
energy levels show a B$^{0.4}$-dependence and therefore a magnetic smearing 
will take place if the flux is integrated over the whole stellar surface. 
For a dipolar field, with a factor of two variation of the magnetic field 
between pole and equator, the spectral features would be broadened by 80 - 
300 eV. Thus it is at least plausible, that the combination of a dense level 
structure of the magnetic atoms with a dispersion of the magnetic field 
produces a spectrum, which appears as a continuum seen with the LETG. We 
believe that this is the most promising model for explaining the featureless 
X-ray spectrum of RX\,J1856. 

Alternatively, the absence of any spectral feature may indicate that the 
star has no atmosphere but a condensed matter surface (Burwitz et al 2001, 
Turolla et al. 2003). Such a surface is expected to be reflective in the 
X-ray domain (Tr\"{u}mper {\&} Lenzen 1978), Brinkmann 1980) which could 
also help to explain the low X-ray/optical flux ratio (see section 3.4). 
Condensation of surface matter requires low temperatures and strong magnetic 
fields. To condense hydrogen at a temperature of kT = 63 eV a magnetic field 
of 5 $\times $ 10$^{13}$ G is required (Lai 2001). For iron it is not clear i
whether a condensate can exist at all. According to Lai (2001) the cohesive 
energy of iron is uncertain, but condensation may possibly occur at 3 
$\times $ 10$^{14}$ G (for kT = 63 eV) while Neuhauser et al. (1987) 
conclude that iron cannot condensate at all. Another problem is that in 
general the optical properties of a condensed matter surface as a function 
of photon energy, polarisation and magnetic field angle have only been 
calculated in the continuum (plasma) approximation while the effects of 
atomic and solid state physics have been neglected. In summary it is not 
clear whether a condensed matter surface can exist and - if it would - 
whether it could provide a solution for the absence of line features.

\subsection{The ``Optical Excess'' and the Absence of Periodic Variations of 
RX\,J1856.5-3754}

Already the ROSAT and optical data had shown that the optical Rayleigh-Jeans 
type spectrum of RX\,J1856 is about a factor of $\sim $3 brighter than the 
extrapolation of the X-ray blackbody towards lower frequencies (e.g. Pons et 
al. 2002). Using current optical data and the LETG spectrum this factor 
turns out to be even larger, namely factor 5-7 (Haberl 2004). This optical 
excess has been explained in terms of an inhomogeneous temperature 
distribution with a hot pole and a cool equator, which would lead to a 
periodic flux variation as observed for four of the seven sources. However, 
for RX\,J1856 the XMM-Newton data put an upper limit of 1.3{\%} on the pulsed 
fraction in the range 50-10$^{-3}$Hz (Burwitz et al. 2003). There are 
several possibilities to explain this behaviour:

\begin{itemize}
\item The rotational frequency could be larger than 50\,Hz, viz. the source 
would be a millisecond pulsar. We regard this case unlikely in view of the 
arguments on its magnetic fields discussed above. Anyway it will be checked 
soon by XMM-Newton EPIC pn observations in the high time resolution mode. 
\item The extreme alternative is, that the neutron star has spun down within 
$\sim$10$^{6}$ years to very long periods, P$>$10000\,sec by the propeller 
effect. This requires an extremely strong magnetic field ($\sim$10$^{15}$\,G) 
and a relatively low velocity (Mori {\&} Ruderman 2003).
\item The simplest explanation is that the rotational axis of the neutron 
star is closely aligned with the line of sight or with the magnetic axis. 
This may look unlikely in view of the low pulsed fraction of $<$1.3\,\%, 
but the average pulsed fraction of the other four sources is only $\sim $12\,\%. 
Therefore the possibility of an accidental alignment cannot be neglected.
\end{itemize}
In this context we note that the tight constraint on the alignment could be 
somewhat relaxed if the X-ray flux were reduced due to reflection effects 
because the size of the X-ray emitting spot would be increased. 

\subsection{A Lower Limit to the Radius of the Neutron Star RX\,J1856.5--3754}

Whatever the answers to the open questions discussed in sections 3.2.--3.4. 
are, one can derive a lower limit for the photospheric radius based on 
blackbody fits for the overall spectrum and on the source distance, as 
discussed by Burwitz et al. (2003) and Tr\"{u}mper et al. (2004). Indeed, 
such a lower limit is expected to be a quite conservative one, keeping in 
mind that a blackbody is the most efficient radiator. With other words: Any 
real emitter needs to have a larger surface than a blackbody radiator in 
order to emit the same luminosity. We stress that the application of this 
rather general ``thermodynamic'' argument seems justified in view of the 
shape of the broadband spectrum, which is characterised by a clear 
Rayleigh-Jeans law in the optical and a Wien-like behaviour at X-ray 
energies. For our analysis we use the distance of 117 pc, which has been 
derived from four HST observations (Walter {\&} Lattimer 2002).

We first consider a simple two-component blackbody model for the optical and 
X-ray spectrum of RX\,J1856 (Burwitz et al. (2003) which is shown in Fig.~4a. 
The blackbody radius and temperature of the X-ray emitting hot spot derived 
from the Chandra LETG spectrum are R$_{x}^{ }$=4.4\,km and kT$_{x}$=63\,eV, 
respectively. The optical spectrum is interpreted as the sum of the 
Rayleigh-Jeans spectra of both the hot and the cool component. This fixes 
(R$_{o}$)$^{2}$$\times$T$_{o}$+(R$_{x}$)$^{2}$$\times$T$_{x}$. The 
condition that the optical spectrum of the cool component does not show up 
as a deviation in the X-ray spectrum limits the corresponding temperature to 
kT$_{0 }<$ 33 eV at the 3$\sigma $ level (Burwitz et al.2003)). Using 
these figures we find for the radius of the neutron star 
R\,=\,(R$_{o}^{2}$\,+\,R$_{x}^{2}$)$^{1/2}$ {\bf $>$ }16.5 km (3$\sigma )$. 

\begin{figure}[htbp]
\centerline{\includegraphics[width=4.60in,height=4.34in]{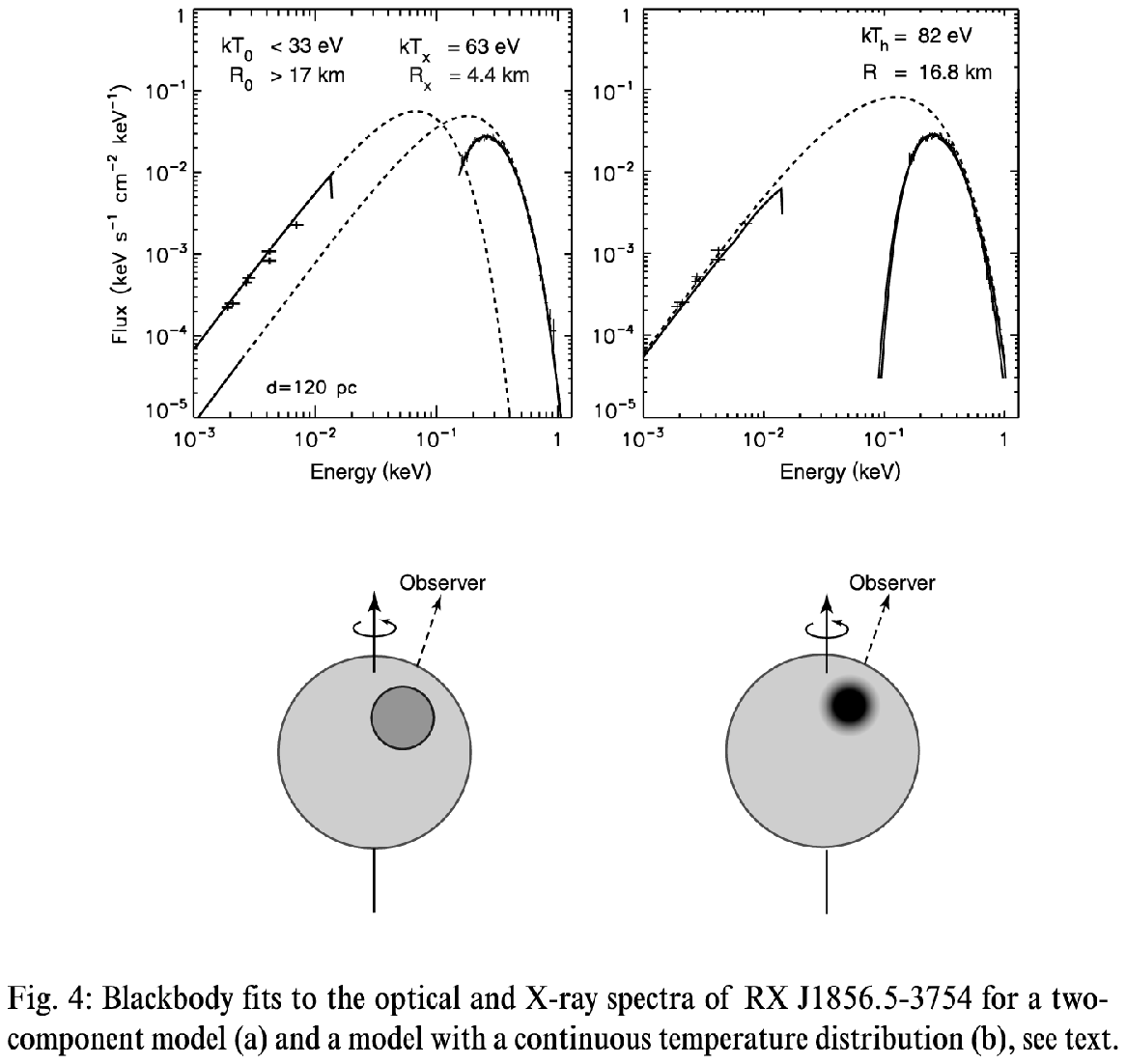}}
\label{fig4}
\end{figure}

As an alternative we use a model with a continuous temperature distribution 
(c.f. Fig.~4b) of the form 

\begin{equation}
T = T_{h} \times  {\{}1+(\theta/\theta _{0})^{\gamma }{\}}^{-1 }
\end{equation}

The best fit to the overall spectrum yields a central temperature of the hot 
spot T$_{h}$\,=\,82\,eV, an angular size of the hot spot $\theta _{0}$\,=\,40$^{0}$ 
and $\gamma$\,=\,2.1. In this case the neutron star radius turns out 
to be 16.8 km ($>$3$\sigma )$, not much different from that of the simpler 
model. 

These apparent radii R measured by a distant observer are related to the 
``true'' stellar radius R$_{0}$ by

\begin{equation}
R = R_{0} (1-R_{s}/R_{0})^{-1/2 }
\end{equation}

where $R_{s}$\,=\,2GM/c$^{2}$ is the Schwarzschild radius. The corresponding 
bound in the M -- $R_{o}$ diagram is shown in Fig.~4. For a standard neutron 
star of 1.4 solar masses the true radii are R$_{0 }$\,=\,14.0\,km (Fig.~4a) and 
R$_{0 }$\,=\,14.1\,km (Fig.~4b), respectively, and thus considerably larger than 
the canonical radius of 10\,km. This implies a rather stiff equation of 
state. We note, that the same conclusion was reached by Braje {\&} Romani 
(2002) using a two-component model and similar arguments. In order to 
compare our results with the predictions of theoretical neutron star models 
in more detail we use the mass-radius diagram given by Pons et al. (2002) 
This diagram is shown in Fig.~5 to which we have added a curve corresponding 
to the apparent radius of R\,=\,16.5\,km.

\begin{figure}[htbp]
\centerline{\includegraphics[width=4.60in,height=3.78in]{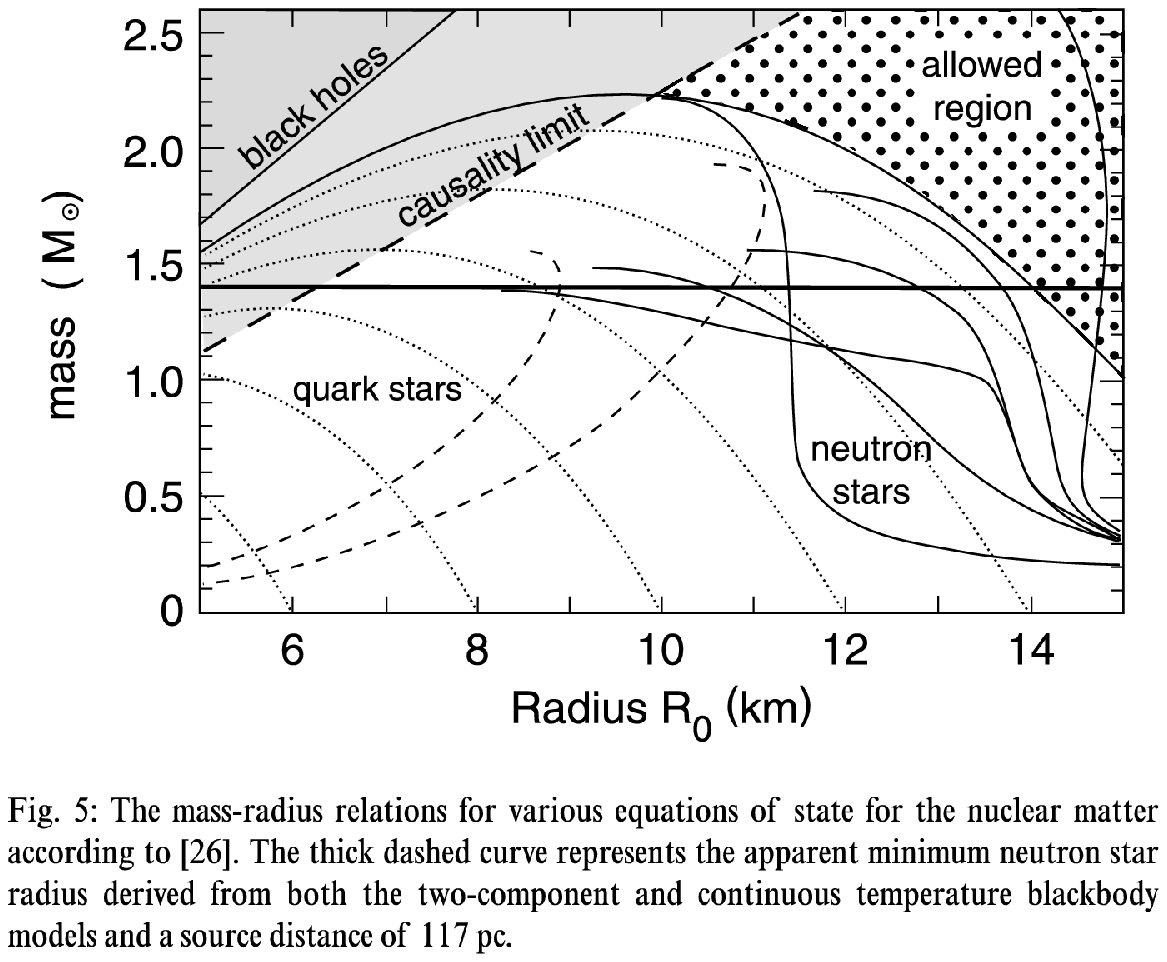}}
\label{fig5}
\end{figure}

It is evident that the result of our analysis excludes the quark star models 
discussed by Pons et al (2002) and by Schertler et al. (1998). Also the 
neutron star models with quark matter cores discussed in the latter paper 
are rejected. We conclude that for a source distance of 117\,pc this neutron 
star must have a very stiff equation of state. Recent improvements of the 
RX\,J1856--3754 parallax, which use additional HST observations at four different 
epochs (Kaplan 2004) yield an even larger distance of 160\,pc. This result 
considerably sharpens our conclusion. In this context one may speculate that 
a too large radius could imply that this neutron star has an anomalously low 
mass ($<$0.4\,M$_{\odot})$. But that would raise a lot of questions; in 
particular it is doubtful whether such a low mass neutron star could be 
formed.

In summary, the observation of RX\,J1856--3754 strongly suggest that the size 
of a neutron star is rather $>$14\,km instead of the canonical size of 10\,km. 
This result gets support from the observations of the three radio 
pulsars discussed in section 2. This has consequences for pulsar 
astrophysics: E.g. the magnetic field strengths estimated from pulsar spin 
down observations have to be lowered by at least a factor of two since B 
$\sim$R$_{o}^{-2}$, and the moment of inertia and therefore the estimate 
of the pulsar rotational energy of a pulsar increases by a similar factor. 
However, the most important result of our analysis is that the behaviour of 
nuclear matter at very high densities is governed by a very stiff equation 
of state.

\section{Outlook}

We have learned a lot about cooling neutron stars from X-ray and optical 
observations since 1990, and we can hope to learn a lot more from Astro-E, 
XEUS and Constellation-X in the future. However, an all-sky survey in the 
soft X-ray band which is at least ten times more sensitive than ROSAT would 
be most important, to find more and fainter sources of this type. At the 
same time, it would be necessary to improve the sensitivity of optical 
observations with instruments of the 30-100 m class in order to measure the 
faint optical spectra. Finally, it seems essential to develop further our 
understanding of atomic and condensed matter physics under the conditions of 
very strong magnetic fields.

\section{Acknowledgements}

The continuous support and cooperation of Werner Becker, Vadim Burwitz, 
Frank Haberl and Slava Zavlin is gratefully acknowledged.

\section{References}

\bibitem{} Becker, W. 2004, private communication.

\bibitem{} Bowyer, C.S., {\it et al.} 1964, Science 146, 912.

\bibitem{} Braje, T.M. {\&} Romani, R.W. 2002, ApJ 580, 1043.

\bibitem{}B rinkmann, W. 1980, A{\&}A 82, 352.

\bibitem{} Brisken, W.F., {\it et al.} 2003, Apj 593, L89.

\bibitem{} Burwitz, V., {\it et al.} 2001, A{\&}A 379, L35.

\bibitem{} Burwitz, V., {\it et al.} 2003, Astron. {\&} Astrophys. 399, 1109.

\bibitem{} Caraveo, P.A., {\it et al.} 1996, ApJ 461, L91.

\bibitem{} Chiu, H.-Y. 1964, Ann. Phys. 2, 364.

\bibitem{}Dodson, R., {\it et al.} 2003, MNRAS 343, 116.

\bibitem{}De Vries, C.P., {\it et al.} 2004, to be published in A{\&}A.

\bibitem{}Haberl, F. 2004, Adv. Space Res. 33, 638.

\bibitem{}Harnden, F.R. {\&} Seward, F.D. 1984, ApJ 283, 279.

\bibitem{}Kaplan, D. 2004, private communication.

\bibitem{}Lai, D. 2001, Rev.Mod.Phys. 73, 629.

\bibitem{}Mc Gowan, K.E., {\it et al.} 2003, ApJ 591, 380.

\bibitem{}Mori, K. {\&} Ruderman, M.A. 2003,ApJ 592, L75.

\bibitem{}Neuhauser, D., {\it et al.} 1987, Phys.Rev. A36, 4163.

\bibitem{}Pavlov, G.G., {\it et al.} 2002, in {\it Proc. of the 270th Haereus Seminar on} {\it Neutron Stars and Supernova Remnants}, W. Becker, H. Lesch, {\&} J. Tr\"{u}mper (eds.) 
MPE Report 278, 273 astro-ph/0206024).

\bibitem{}Pavlov, G.G. {\&} Zavlin, V.E. 2003, in {\it Proceedings of the XXI Texas Symposium on relativistic Astrophysics}, B. Rino, R. Maiolino, {\&} M. 
Filippo (eds.), astro-ph/ 0305435).

\bibitem{}Pavlov, G.G., {\it et al.} 1996, ApJ 472, L33.

\bibitem{}Pons, J.A., {\it et al.} 2002, ApJ 564, 981.

\bibitem{}Rajagopal, M., {\it et al.} 1997, ApJ 479, 347.

\bibitem{}Schertler, K., {\it et al.} 1998, Nucl. Phys. A637, 451.

\bibitem{}Tr\"{u}mper, J. {\&} Lenzen, R. 1978, Nat 271, 216.

\bibitem{}Tr\"{u}mper, J. 1983, Adv. Space Res. 2, 142.

\bibitem{}Tr\"{u}mper, J., {\it et al.} 2004, Nucl. Phys., Vol 132C, 560.

\bibitem{}Turolla, R., {\it et al.} 2003, ApJ 603, 265.

\bibitem{}Van Kerkwijk, M.H. {\&} Kulkarni, S.R. 2001, A{\&}A 380, 221.

\bibitem{}Walter, F.M., {\it et al.} 1996, Nat 379, 233.

\bibitem{} Walter, F.M. {\&} Lattimer, J. 2002, ApJ 576, L145.

\bibitem{} Walter, F.M. {\&} Matthews, L.D. 1997, Nat 389, 358.

\bibitem{} Weisskopf, M. 2004, (this volume).

\bibitem{} Zavlin, V.E. {\&} Pavlov, G.G. 2002, in {\it Proc. of the 270}$^{th}${\it  Haereus Seminar on Neutron Stars, Pulsars and Supernova Remnants}, W. Becker, H. Lesch {\&} J. 
Tr\"{u}mper (eds.), MPE Report 278, 263.

\end{document}